\newif\ifpreprint
 
\preprintfalse 
  
\ifpreprint
\documentclass[journal=jctcce,manuscript=article]{achemso}
\else
\documentclass[journal=jctcce,manuscript=article,layout=twocolumn]{achemso}
\fi

\usepackage[T1]{fontenc} 

\usepackage{amsmath}
\usepackage{newtxtext,newtxmath}

\usepackage{graphicx}
\usepackage{dcolumn}
\usepackage{braket}
\usepackage{multirow}
\usepackage{threeparttable}
\usepackage{xspace}
\usepackage{verbatim}
\usepackage[version=4]{mhchem} 
\usepackage{comment}
\usepackage{color,soul}

\usepackage{mathtools}
\usepackage{xcolor}
\usepackage{xspace}
\usepackage{ifthen}

\usepackage{qcircuit}

\newcommand*{\Eh}{$E_{\rm h}$\xspace}

\newcommand{\basis}[2]{\psi_{#1}^{({#2})}\xspace}
\newcommand{\familyabbr}[0]{QSD\xspace}
\newcommand{\familyname}[0]{Quantum Subspace Diagonalization\xspace}
\newcommand{\methodname}[0]{multireference selected quantum Krylov\xspace}
\newcommand{\methodabbr}[0]{MRSQK\xspace}
\newcommand{\controlop}[1]{\mathrm{c}\mhyphen{#1}}
\mathchardef\mhyphen="2D

\usepackage[colorlinks = true,
            linkcolor = blue,
            urlcolor  = black,
            citecolor = blue,
            anchorcolor = black]{hyperref}

\definecolor{goodorange}{RGB}{225,125,0}
\definecolor{goodgreen}{RGB}{5,130,5}
\definecolor{goodred}{RGB}{220,50,25}
\definecolor{goodblue}{RGB}{30,144,255}
\definecolor{OliveGreen}{RGB}{5,100,5}

\newcommand{\note}[2]{
\ifthenelse{\equal{#1}{F}}{
\colorbox{goodorange}{\textcolor{white}{\footnotesize \fontfamily{phv}\selectfont #1}}
    \textcolor{goodorange}{{\footnotesize \fontfamily{phv}\selectfont #2}}\xspace
}{}
\ifthenelse{\equal{#1}{R}}{
\colorbox{goodred}{\textcolor{white}{\footnotesize \fontfamily{phv}\selectfont #1}}
    \textcolor{goodred}{{\footnotesize \fontfamily{phv}\selectfont #2}}\xspace
}{}
\ifthenelse{\equal{#1}{N}}{
\colorbox{goodgreen}{\textcolor{white}{\footnotesize \fontfamily{phv}\selectfont #1}}
    \textcolor{goodgreen}{{\footnotesize \fontfamily{phv}\selectfont #2}}\xspace
}{}
\ifthenelse{\equal{#1}{M}}{
\colorbox{goodblue}{\textcolor{white}{\footnotesize \fontfamily{phv}\selectfont #1}}
    \textcolor{goodblue}{{\footnotesize \fontfamily{phv}\selectfont #2}}\xspace
}{}
}

\usepackage{titlesec}

\usepackage[fontsize=11pt]{scrextend}
\titleformat{\section}
{\normalfont\sffamily\bfseries\color{OliveGreen}}
{\thesection.}{0.25em}{\uppercase}

\titleformat{\subsection}[runin]
{\normalfont\sffamily\bfseries}
{\thesubsection}{0.25em}{}[.\;\;]

\titleformat{\suppinfo}
{\normalfont\sffamily\bfseries}
{\thesubsection}{0.25em}{}

\titlespacing*{\section}{0pt}{0.5\baselineskip}{0.01\baselineskip}
\titlespacing*{\subsection}{0pt}{0.125\baselineskip}{0.01\baselineskip}

\author{Nicholas H. Stair}
\affiliation{Department of Chemistry and Cherry Emerson Center for Scientific Computation, Emory University, Atlanta, GA, 30322}

\author{Renke Huang}
\affiliation{Department of Chemistry and Cherry Emerson Center for Scientific Computation, Emory University, Atlanta, GA, 30322}

\author{Francesco A. Evangelista}
\email{francesco.evangelista@emory.edu}
\affiliation{Department of Chemistry and Cherry Emerson Center for Scientific Computation, Emory University, Atlanta, GA, 30322}

\let\oldmaketitle\maketitle
\let\maketitle\relax

\title{A Multireference Quantum Krylov Algorithm for Strongly Correlated Electrons}

\date{\today}

\begin{document}

\ifpreprint
\else
\twocolumn[
\begin{@twocolumnfalse}
\fi
\oldmaketitle

\begin{abstract}

We introduce a \methodname (\methodabbr) algorithm suitable for quantum simulation of many-body problems.
\methodabbr is a low-cost alternative to the quantum phase estimation algorithm that generates a target state as a linear combination of non-orthogonal  Krylov basis states. This basis is constructed from a set of reference states via real-time evolution avoiding the numerical optimization of parameters.
An efficient algorithm for the evaluation of the off-diagonal matrix elements of the overlap and Hamiltonian matrices is discussed and a selection procedure is introduced to identify a basis of orthogonal references that ameliorates the linear dependency problem.
Preliminary benchmarks on linear \ce{H6}, \ce{H8}, and \ce{BeH2} indicate that \methodabbr can predict the energy of these systems accurately using very compact Krylov bases.
\end{abstract}

\ifpreprint
\else
\end{@twocolumnfalse}
]
\fi

\ifpreprint
\else
\small
\fi

\noindent


\section{Introduction}

Solving the electronic many-body Schr\"{o}dinger equation for systems that display strong correlation effects is a major challenge in physics and quantum chemistry.\cite{Laughlin:2000br}
Quantum computation\cite{Feynman:1982gn} offers a potential solution to the exponential scaling of the Hilbert space dimension with particle number.
Recent advances in quantum hardware design, including an early demonstrations of quantum speedup,\cite{Arute:2019fg} have motivated the development of new quantum algorithms that can be executed on so called  noisy intermediate-scale quantum (NISQ) devices, with less than 100 qubits and shallow circuits.\cite{Preskill:2018gt}

Algorithms based on quantum phase estimation (QPE),\cite{Abrams:1997ha,Abrams:1999ur} were the first proposed to compute the ground state energies of fermionic many-body systems.\cite{ortiz2001quantum}
QPE was later applied to molecular problems\cite{AspuruGuzik:dj} and has been implemented on a photonic quantum device.\cite{lanyon2010towards}
Though QPE is well suited for Hamiltonian simulation on large-scale fault-tolerant quantum hardware, its application in the NISQ era presents several challenges due to the poor gate fidelity and the limited coherence time of devices available in the foreseeable future.\cite{McArdle:2018we,Cao:2019jn}
As a result, hybrid quantum-classical algorithms requiring shallower circuits, such as the variational quantum eigensolver (VQE)\cite{Peruzzo:2014kca, yung:2014} and the quantum approximate optimization algorithm (QAOA)\cite{Farhi:2014wl} have recently received more attention.

In the VQE scheme, a complex trial wave function is optimized via an algorithm that subdivides the work between a classical and quantum computer. In this approach, the variational minimization of the energy is driven by a classical algorithm, while measurement of the energy and gradients is deployed to a quantum computer.
VQE was originally implemented with the unitary coupled cluster (UCC)\cite{theory:1977bs, Szalay:1995vu,Taube:2006uy, Cooper:2010ck, Evangelista:2011eh,Harsha:2018dv} ansatz truncated to single and double excitations.\cite{Peruzzo:2014kca, yung:2014, McClean:2016bs, OMalley:2016dc, Romero:2019hk, Barkoutsos:2018hm} 
More recently, several groups have studied alternative ans\"{a}tze, including mean-field references,\cite{Ryabinkin:2018hd} UCC with general singles and doubles,\cite{Wecker:2015da} hardware-efficient parameterizations,\cite{Kandala:2017gh} resource-efficient qubit-space UCC with 2-qubit entanglers,\cite{Ryabinkin:2018uc, Ryabinkin:2019vh} general UCC with adaptively selected unitaries,\cite{Grimsley:2019ed} and linear-depth fermionic Gaussian reference states.\cite{DallaireDemers:2019iw}
Efforts have also been made to extend the VQE algorithm to compute excited states\cite{McClean:2017ct, Colless:2018hp, Higgott:2019fc, Nakanishi:2018wo, Jouzdani:2019tp, Parrish:2019bw} and approaches that combine variational methods and phase estimation have been suggested.\cite{Santagati:2018ih, Wang:2019ha}

Notwithstanding the significant impact of VQE schemes, they have two principal drawbacks.
Firstly, VQE methods require measurement of the energy or energy gradients with respect to the variational parameters at each step of the optimization process.
This results in a significant number of queries of the optimization algorithm to the quantum device.
Secondly, the optimization process in VQE is challenging due to the high nonlinearity of the energy (considered as a function of the parameters), intrinsic accuracy limitation because of the inexactness of the ansatz\cite{evangelista2019exact} and stochastic errors that result from finite measurement and loss of fidelity.\cite{Barkoutsos:2019wq}
As a consequence, the optimization process may be slow to convergence and may reach a local minimum instead of the true ground state.

A third and emerging family of methods, which we refer to as \familyname (\familyabbr) schemes, diagonalize the Hamiltonian in a general nonorthogonal basis of many-body states.\cite{
McClean:2017ct, Colless:2018hp, Takeshita:2019wn,Motta:2019ul,Huggins:2019vv,Parrish:2019tc}
There is a long tradition of using such a strategy in quantum chemistry.\cite{Lowdin:1950tq,King:1967ev,Noodleman:1981iz,Voter:1981eu,Malmqvist:1986jh,Koch:1993tf}
A natural way to extend it to quantum computing is to construct a basis of states and measure the corresponding matrix elements with a quantum device, and later solve the associated generalized eigenvalue problem via a classical computer.\cite{McClean:2017ct}
Compared to a fully classical approach, \familyabbr schemes can take advantage of the ability of quantum computers to store arbitrarily complex states.

\familyabbr methods mainly differ in the way the many-body basis is generated.
The quantum subspace expansion (QSE) method of McClean and co-workers, diagonalizes the Hamiltonian in the basis of states $\hat{a}_{i}^\dagger\hat{a}_{j}\ket{\Psi}$, where $\Psi$ is a reference state prepared via VQE.\cite{McClean:2017ct, Colless:2018hp, Takeshita:2019wn}
Matrix element of the Hamiltonian in this basis are obtained by measuring the three- and four-body density matrices.
\familyabbr approaches are particularly advantageous if the many-body basis is constructed as a Krylov space\cite{Motta:2019ul} and does not require extensive parameter optimization.
This is the case for the Quantum Lanczos (QLanczos) algorithm,\cite{Motta:2019ul}where the Hamiltonian is diagonalized in a basis of correlated states generated by imaginary-time propagation.\cite{McArdle:2019ek} This basis is obtained from a single reference state by sampling at regular intervals in imaginary time. In QLanczos, the imaginary-time propagator is written as a unitary operation times a normalization factor, and a linear approximation is employed to construct this representation.
For each step in the imaginary-time propagation, a linear system of equations must be solved for classically, requiring the measurement of the entries of a matrix and a vector.
Recently, Huggins \textit{et al.}\cite{Huggins:2019vv} have proposed a nonorthgonal VQE (NOVQE) scheme whereby the basis is formed of $k$-fold products of unitary paired coupled cluster with generalized single and double excitations ($k$-UpCCGSD).\cite{Lee:2018cy}
Parrish and McMahon,\cite{Parrish:2019tc} investigated a quantum filter diagonalization (QFD) formalism in which a basis of states is generated via an approximate real-time dynamics.
QFD is inspired by classical filter diagonalization\cite{neuhauser1990bound, neuhauser1994circumventing, wall1995extraction, mandelshtam1997low} as well as quantum time grid methods,\cite{somma2002simulating, o2019quantum, kyriienko2019quantum, somma2019quantum} and may be viewed as a variant of the QLanczos algorithm in which imaginary time propagation is replaced with a real-time version.

Despite their potential, \familyabbr methods suffer from a series of practical issues, which are the focus of this work.
The generalized eigenvalue problem associated with a given nonorthogonal basis requires the efficient evaluation of off-diagonal matrix elements of the form $\braket{\psi _\alpha|\hat{O}| \psi _\beta}$.
While in the case of QSE and QLanczos these matrix elements are easily computed,\cite{McClean:2017ct,Motta:2019ul} in the general case their evaluation is more involved.\cite{Parrish:2019tc,Huggins:2019vv}
Another important issue is the linear dependency of the basis generated in a \familyabbr procedure.
This issue introduces numerical instabilities in the generalized eigenvalue problem and is potentially amplified by poor gate fidelity and measurement errors.
Bases generated by variational optimization\cite{Huggins:2019vv} and real\cite{Parrish:2019tc} or imaginary\cite{Motta:2019ul} time propagation are all plagued (to various degrees) by linear dependencies.

In this work we formulate a \familyabbr algorithm that addresses the two problems described above. Firstly, we describe an efficient approach to evaluate the off-diagonal matrix element required in \familyabbr methods, with a cost that is essentially identical to that of computing a diagonal matrix element.
Secondly, to mitigate the linear dependency problem we consider a multireference approach in which the Krylov space is constructed from an initial set of orthogonal reference states. 
These references are selected via a scheme that exploits quantum measurement to identify the most important determinants in a simple trial wave function.
The resulting \methodname (\methodabbr) method is combined with basis generation via real-time propagation\cite{Motta:2019ul,Parrish:2019tc} and  benchmarked on a series of problems involving strongly correlated electronic states.

\section{Theory}

Consider a molecular Hamiltonian mapped to a set of qubits ($\hat{H}$)
\begin{equation}
\hat{H} = E_0 + \sum_\ell h_\ell \hat{V}_\ell
\end{equation}
where $E_0$ is a scalar term, the index $\ell$ runs over all the terms in the Hamiltonian, $h_\ell$ is a matrix element, and $\hat{V}_\ell$ is the corresponding operator.
Each operator $\hat{V}_\ell$ in $\hat{H}$ is a tensor product of $N_{\ell}$ Pauli operators (a Pauli string) that act on distinct qubits,
$\hat{V}_\ell =  \bigotimes_{k=1}^{N_{\ell}} \sigma_{l_k}^{(j_k)},$
where $l_k \in \{ X, Y, Z\}$ labels the Pauli operator type and $j_k$ indicates the qubit upon which said operator is applied.

To define the \methodabbr method, we start by introducing a $d$-dimensional basis of reference states, $\mathcal{M}_0 = \{ \Phi_I \}$, where each $\Phi_I$ is a linear combination of Slater determinants ($\phi_\mu$) with well defined spin and spatial symmetry
\begin{equation}
\ket{\Phi_I} = \sum_\mu d_{\mu I} \ket{\phi_\mu} 
\end{equation}
From this basis, we generate a nonorthogonal Krylov\cite{Saad:2003gc,Furche:2016bn} space $\mathcal{K}_s(\mathcal{M}_0, \hat{U}_n) = \{ \psi_\alpha, \alpha = 1, \ldots, N \}$ by repeated application of a family of unitary operators $\hat{U}_n$ (with $n = 0, 1, \ldots, s$) to all the elements of $\mathcal{M}_0$.
A generic element $\basis{I}{n} \in \mathcal{K}$ is given by the action of $\hat{U}_n$ on $\Phi_I.$
\begin{equation}
 \ket{\psi_\alpha} \equiv  \ket{\basis{I}{n}} = \hat{U}_n \ket{\Phi_I}
\end{equation}
For convenience, we use the collective index $\alpha = (I,n)$ to identify an element of the basis.
The resulting Krylov space has dimension $N = d (s+1)$.

In \methodabbr, a general state is written as a linear combination of the basis $\{ \psi_\alpha \}$ as
\begin{equation}
\ket{\Psi} = \sum_\alpha  c_\alpha \ket{\psi_\alpha}= \sum_{I=1}^{d} \sum_{n=0}^{s} c_{I}^{(n)} \hat{U}_n \ket{\Phi_I} 
\label{eq:wfn}
\end{equation}
Variational minimization of the energy of the state $\Psi$ leads to the following generalized eigenvalue problem
\begin{equation}
\mathbf{Hc} = \mathbf{Sc} E,
\label{eq:gep}
\end{equation}
where the elements of the overlap matrix ($\mathbf{S}$) and Hamiltonian ($\mathbf{H}$) are given by
\begin{align}
\label{eq:overlap}
S_{\alpha\beta} &= \braket{\psi_\alpha | \psi_\beta}
= \braket{\Phi_{I} | \hat{U}^\dagger_m \hat{U}_n |\Phi_J},\\
\label{eq:hamiltonian}
H_{\alpha\beta} &= \braket{\psi_\alpha | \hat{H} | \psi_\beta}
= \braket{\Phi_{I} | \hat{U}^\dagger_m \hat{H} \hat{U}_n |\Phi_{J}}
\end{align}
The formalism outlined above lends itself to a large number of quantum algorithms, depending on: i) how the basis $\mathcal{M}_0$ is selected, ii) the particular choice of $\hat{U}_n$, and iii) the quantum circuits used to evaluate $\mathbf{S}$ and $\mathbf{H}$.
In the following we describe the combination that defines our \methodname approach and detail the efficient algorithm used to evaluate off-diagonal overlap and Hamiltonian matrix elements and our selection approach to generate the basis of references.

\begin{figure}[h!]
\centering
\includegraphics[width=3.25in]{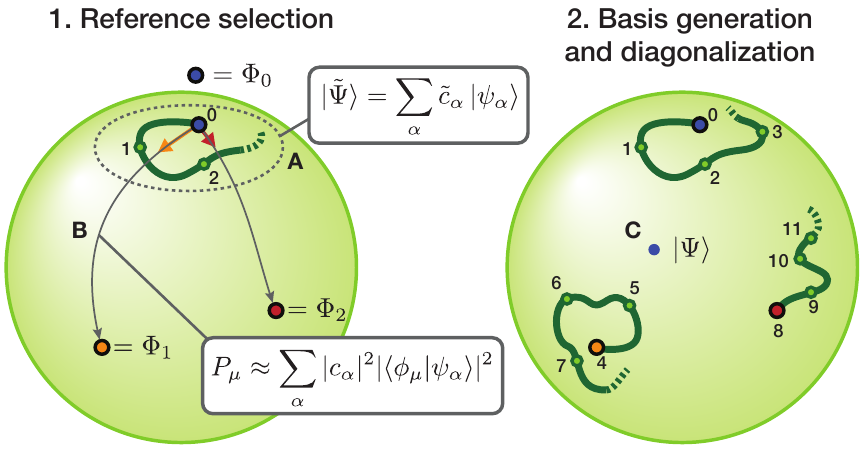}
\caption{Schematic illustration of the \methodname (\methodabbr) algorithm. (A) An approximate real-time dynamics using a single Slater determinant reference ($\Phi_0$) is used to generate a trial state ($\tilde{\Psi}$). 
(B) Measurements of the determinants that comprise the trial state are used to determine the probability of hopping ($P_\mu$) to other determinants. This information is employed to build two new reference states, $\Phi_1$ and $\Phi_2$. (C) Finally, three real-time evolutions starting from the references ($\Phi_0, \Phi_1, \Phi_2$) generate a set of 12 Krylov states $\psi_\alpha$, which are used to diagonalize the Hamiltonian and obtain the energy of the state $\Psi$.}
\label{fig:H6pes_5e_1_exst}
\end{figure}

\subsection{Choice of the unitary operators}
\label{sec:choice_of_unitary}

In choosing the family of unitary operators  $\hat{U}_n$ there are two  primary criteria we aim to satisfy: i) that it generates a basis that well describes the eigenstates of $\hat{H}$ and ii) that the corresponding quantum circuit is inexpensive to evaluate.
These requirements give considerable freedom, and a natural choice is a family of operators based on real-time evolution, $\hat{U}_n = \exp(-i t_n \hat{H})$, where $t_n = n \Delta t$ and $\Delta t$ is a fixed time step.

In fact, it is possible to show that for small $\Delta t$, the basis of states $\mathcal{K}_s(\Phi_I, \hat{U}_n)$ generated by real-time evolution spans a classical Krylov space.
Consider a linear combination of the elements of $\mathcal{K}_s(\Phi_I, \hat{U}_n)$ and expand the exponential into a Taylor series keeping terms up to order $(\Delta t)^{s}$
\begin{equation}
\begin{split}
\ket{\Psi} = & \sum_{n=0}^s c_{I}^{(n)} e^{-i n \Delta t \hat{H}} \ket{\Phi_I} \\
= & \sum_{k=0}^s \bigg( \sum_{n=0}^s \frac{(-i n \Delta t)^k}{k !} c_{I}^{(n)} \bigg) \hat{H}^k \ket{\Phi_I} + \mathcal{O}(\Delta t^{s+1}) \\
=& \sum_{k=0}^s \bigg( \sum_{n=0}^s M_{kn} c_{I}^{(n)} \bigg) \hat{H}^k \ket{\Phi_I} + \mathcal{O}(\Delta t^{s+1})
\end{split}
\end{equation}
The square matrix $\mathbf{M}$ is invertible, and therefore, the coefficients $c_{I}^{(n)}$ may be chosen to represent any combination of the classical Krylov basis $\{\hat{H}^k \ket{\Phi_I}\}$ with $k = 0, \ldots, s$, up to higher-order terms.

To realize the \methodabbr method on a quantum computer, a quantum circuit is required that can approximate the time-evolution operator.
The analysis presented above, suggests that it is important that any approximation to the real-time evolution operator must be sufficiently accurate. Otherwise the approximate quantum Krylov basis will likely not span the classical Krylov basis of the exact Hamiltonian, and consequently slow down the convergence of the method.
To approximate the real-time evolution one may follow standard approaches like the Trotter-Suzuki decomposition\cite{Suzuki:1977bw,Lloyd:1996ch} or employ a truncated Taylor series.\cite{Berry:2015ir}
In this work we employ the former methodology, and consider the $m$-Trotter number (step) approximation of non-commuting operators $\hat{A}$ and $\hat{B}$ given by
\begin{equation}
e^{\hat{A} + \hat{B}} \approx \big(  e^{ \frac{ \hat{A}}{m} } e^{\frac{ \hat{B}}{m}}  \big)^{m}
\label{eq:m_trot_step}
\end{equation}
which is exact in the limit of $m \rightarrow \infty$.
When applied to the real-time propagator this corresponds to the product
\begin{equation}
\label{eq:trotter_ham}
\hat{U}_n = \left( \prod_\ell \hat{U}_{n, \ell}(t_n/m)\right)^m = \left( \prod_\ell \exp(-i t_n h_\ell / m \hat{V}_\ell) \right)^m
\end{equation}
As shown in section~\ref{sec:results}, low Trotter number approximations ($m = 1,2$) yield large errors in the computation of the ground state electronic energy.

\subsection{Efficient evaluation of off-diagonal matrix elements}
\label{sec:ev_off_diag_elem}
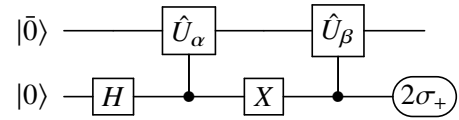
\begin{figure}[bh!]
\[ \Qcircuit @C=1em @R=.7em {
   & \lstick{\ket{\bar{0}}}  &\qw                    & \gate{\hat{U}_\alpha} &   \qw          & \gate{\hat{U}_\beta} & \qw \\
   & \lstick{\ket{0}}  &\gate{H}     & \ctrl{-1} \qwx        &   \gate{X}   & \ctrl{-1} \qwx             & \measure{\mbox{$2\sigma_{+}$}} \\
} \]
\caption{General circuit for measuring non-hermitian operators of the form $\bra{\bar{0}} \hat{U}_\alpha^\dagger \hat{U}_\beta \ket{\bar{0}} $. In this circuit, the final measurement corresponds to separate measurements of $X$ and $Y$ and the evaluation of the expectation value of the operator $ 2\sigma_+ = X  + i  Y  = 2\ket{0}\bra{1}$.}
\label{fig:gen_overlap_circ}
\end{figure}

To efficiently measure the overlap and Hamiltonian matrix elements [Eqs.~\ref{eq:overlap} and~\ref{eq:hamiltonian}], we augment the circuit used to build the basis with an ancillary qubit and construct the state $\frac{1}{\sqrt{2}} ( \ket{\psi_\alpha}\otimes \ket{0}+  \ket{\psi_\beta}\otimes \ket{1} )$, and then obtain $\braket{\psi_\alpha| \psi_\beta}$ by measuring the expectation value of the operator $2\sigma_{+} = \sigma_X + i \sigma_Y$ on the ancilla qubit.\cite{somma2003quantum}
To produce the state $\ket{\psi_\alpha}$ we introduce the unitary operator  $\hat{U}_\alpha$ defined as
\begin{equation}
\hat{U}_\alpha = \hat{U}_n \hat{U}_I
\end{equation}
where $\hat{U}_I$ generates the reference state $\ket{\Phi_I}$ from the zero state $\ket{\bar{0}} = \prod \ket{0}$.
The circuit to measure off-diagonal matrix elements is shown in Fig.~\ref{fig:gen_overlap_circ}.

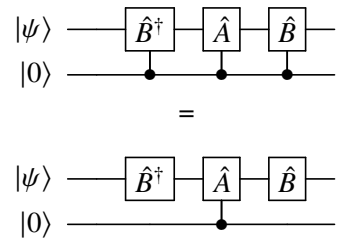
\begin{figure}[b!]
\[ \Qcircuit @C=1em @R=.7em {
   & \lstick{\ket{\psi}}  &\qw       & \gate{\hat{B}^\dagger}     &   \gate{\hat{A}}         & \gate{\hat{B}} & \qw \\
   & \lstick{\ket{0}}      &\qw       & \ctrl{-1} \qwx                      &   \ctrl{-1} \qwx  & \ctrl{-1} \qwx       &   \qw  \\
} \]
=
\[ \Qcircuit @C=1em @R=.7em {
   & \lstick{\ket{\psi}}  &\qw       & \gate{\hat{B}^\dagger}     &   \gate{\hat{A}}         & \gate{\hat{B}} & \qw \\
   & \lstick{\ket{0}}      &\qw       & \qw                                    &   \ctrl{-1} \qwx  & \qw                   &   \qw  \\
} \]
\caption{Circuit identity used to simplify the controlled version of $\hat{U}_{n, \ell}$ [Eq.~\eqref{eq:op_identity}]. $\psi$ is a multi-qubit register used to encode a quantum state and the last qubit is an ancilla.}
\label{fig:controll_eq_circ}
\end{figure}

For $\hat{U}_n$ constructed out of exponentials of Pauli strings a crucial simplification may be employed that allows the efficient construction of the state $\frac{1}{\sqrt{2}} ( \ket{\psi_\alpha}\otimes \ket{0}+  \ket{\psi_\beta}\otimes \ket{1} )$.\cite{Hastings:2015}
First, we start by representing the product of Pauli operators in each of terms $\hat{V}_\ell$ as a unitarily-transformed Pauli string consisting of operators in the $Z$ basis
\begin{equation}
\hat{V}_\ell = \bigotimes_{k=1}^{N_{\ell}} \sigma_{l_k}^{(j_k)}
= \mathcal{H}_\ell \bigotimes_{k=1}^{N_{\ell}} \sigma_{Z}^{(j_k)} \mathcal{H}_\ell
\end{equation}
where $\mathcal{H}_\ell$ is a product of single qubit gates that transform each $\sigma_{l_k}^{(j_k)}$ to $\sigma_{Z}^{(j_k)}$.\cite{Seeley:2012em}
Consequently, each term in $\hat{U}_{n, \ell} = \exp(-i t_n h_\ell \hat{V}_\ell)$ can be written as
\begin{equation}
\label{eq:Unl_exp}
\hat{U}_{n, \ell} =
\mathcal{H}_\ell
\big( e^{-i t_n h_\ell\bigotimes_{k=1}^{N_{\ell}} \sigma_{Z}^{(j_k)}} \big)
\mathcal{H}_\ell
=  \tilde{U}_{n, \ell} R_{z_{N_{\ell}}}(2 t_n h_\ell) \tilde{U}_{n, \ell}
\end{equation}
where in the second step we have used a well-known representation of the exponential of Pauli strings composed of CNOT gates (collected in $\tilde{U}_{n, \ell}$) and a Z rotation on the $N_\ell$ qubit ($R_{z_{N_{\ell}}}$).\cite{somma2003quantum, Seeley:2012em}
Using the following operator identity involving the controlled versions of generic unitary operators $\hat{A}$ and $\hat{B}$ ($\controlop{\hat{A}}$ and $\controlop{\hat{B}}$, see Fig.~\ref{fig:controll_eq_circ}).
\begin{equation}
\label{eq:op_identity}
(\controlop{\hat{B}^\dagger}) (\controlop{\hat{A}}) (\controlop{\hat{B}})=
\hat{B}^\dagger (\controlop{\hat{A}}) \hat{B}
\end{equation}
We can rewrite the controlled unitary evolution operator ($\controlop{\hat{U}_{n, \ell}}$) required to evaluate overlaps as
\begin{equation}
\label{eq:Unl_exp_simple}
\controlop{\hat{U}_{n, \ell}} = \tilde{U}_{n, \ell}^\dagger [ \controlop{R_{z_{N_{\ell}}}(2 \theta_n)} ]\tilde{U}_{n, \ell}
\end{equation}
which requires only one extra controlled operation $\controlop{R_{z_{N_{\ell}}}(2 \theta_n)}$ at the center of the circuit (see Supporting Information Fig.~SI1).
Controlled unitaries evaluated in this way require at most $2 N_\ell$ single qubit gates, $2N_\ell$ CNOT gates, and a controlled single-qubit gate.

Next, we discuss the the implementation of the unitary that prepares reference states from $\ket{\bar{0}}$ ($\hat{U}_I$). When $\Phi_I$ is a single Slater determinant $\hat{U}_I$ is a product of $X$ gates.
For multideterminantal references, one can apply the linear combination of unitaries (LCU) algorithm,\cite{Childs:2012cg} or follow the procedure outlined by Tubman \textit{et al.}\cite{Tubman2018postponing}. This approach requires only one ancilla qubit and $\mathcal{O}(nL)$ one- or two-qubit gates, where $n$ is the number of qubits and $L$ the number of determinants in a particular reference
Alternatively, one may target references that are composed of a single configuration state functions\cite{sugisaki2016quantum, sugisaki2019open} or two electron geminals.\cite{Sugisaki:2019fz}

It is easy to generalize these circuits to controlled versions; however, one may pay the penalty of increasing the number of two-qubit gates (after factoring three qubit control gates into two-qubit ones).
This suggests that the references $\Phi_I$ should be chosen to be compact multideterminantal wave functions, e.g., either single determinants or a small linear combinations of determinants.

Evaluation of the Hamiltonian matrix elements $H_{\alpha\beta}=\sum_\ell h_\ell \braket{\psi_\alpha|\hat{V}_\ell|\psi_\beta}$ proceeds in an analogous way by computing each term $\braket{\psi_\alpha|\hat{V}_\ell|\psi_\beta}$ individually.
The circuit employed is analogous to the one in Fig.~\ref{fig:gen_overlap_circ} with the operator $\hat{U}_\beta$   by replaced by $\hat{V}_\ell\hat{U}_\beta$.
Since each term $\hat{V}_\ell$ contains only product of one qubit operators, the corresponding controlled operator contains at most two-qubit operators.
The evaluation of $\mathbf{S}$ and $\mathbf{H}$ lends itself to a high degree of parallelism.
As in VQE methods, evaluation of a single matrix element of $\mathbf{H}$ may be parallelized over terms in the Hamiltonian.
In addition, in the \methodabbr, one may parallelize over the $N(N-1)/2$ unique pairs of Krylov states $\psi_\alpha$/$\psi_\beta$.
Note that techniques used to ameliorate finite measurement errors in VQE\cite{izmaylov2019revising, izmaylov2019unitary, gokhale2019minimizing, zhao2019measurement, gokhale2019n} approaches can also be applied to \methodabbr.

\subsection{Reference selection}
A third important aspect of the \methodabbr algorithm is the procedure to select the reference configurations.
Our approach exploits quantum measurement to identify a set of configurations starting from a trial \methodabbr wave function.
Specifically, we first form and diagonalize the Hamiltonian in the Krylov space $\mathcal{K}_s(\Phi_0, \hat{U}_n)$, where $\Phi_0$ is a single determinant (e.g., a closed-shell Hartree--Fock determinant).
The resulting trial wave function $\tilde{\Psi} = \sum_\alpha \psi_\alpha \tilde{c}_\alpha$ is used to construct a list of potential important determinants.
Since the probability of measuring a determinant $\phi_\mu$ is equal to $P_\mu = |\langle\phi_\mu|\tilde{\Psi}\rangle|^2$, one can in principle form the state $\tilde{\Psi}$ on a quantum computer and directly measure the determinantal composition, which in the Jordan--Wigner mapping amounts to measuring the expectation value of $Z$ for all wave function qubits.
In practice, we approximate $P_\mu $ by measuring each element of the Krylov basis and estimating the total probability as a weighted sum over references via
\begin{equation}
P_\mu = |\sum_\alpha \braket{\phi_\mu|\psi_\alpha} c_\alpha|^2
\approx
\sum_\alpha |\braket{\phi_\mu|\psi_\alpha}|^2 |c_\alpha|^2
\label{eq:etaMA}
\end{equation}
Measurements are accumulated until we form a list of determinants of length equal to a small multiple of the number of references we aim to select (e.g., $2d$).
In principle only a small number of measurements are required because the values of $P_\mu$ need only be qualitatively correct such that the determinants can be sorted.
It should be noted, however, that using Eq.~\eqref{eq:etaMA} as an importance criterion can lead to the overestimation of the importance of certain determinants due to neglected sign cancellation.
A comparison of the the approximate sampling based on Eq.~\eqref{eq:etaMA} and the exact weight of determinants in the \methodabbr wave function shows that the former method is sufficient to identify the most important determinants.
Alternatively, $\Psi$ could be directly represented on a quantum computer via the linear combination of unitaries (LCU) algorithm,\cite{Childs:2012cg} so that determinants would be sampled directly with their correct probabilities.

Once formed, the list of potentially important determinants is augmented to guarantee that all spin arrangements of open-shell determinants are included.
Next, we diagonalize the Hamiltonian in this small determinant basis. At this stage we identify references in the following way: closed-shell determinants are considered individually, while open-shell determinants with the same spin occupation pattern are grouped together and their weight summed.
Lastly, we select $d-1$ largest weighted references beyond the Hartree--Fock state. References composed of open-shell determinants are normalized to one using the determinant coefficients from the small classical CI.
This procedure generates very compact reference states that can be used with the algorithm for computing off-diagonal matrix elements discussed in section~\ref{sec:ev_off_diag_elem}.

\subsection{Analysis of computational cost}
The quantum computational cost of the \methodabbr algorithm is dominated by the application of the Trotterized Hamiltonian circuits $\hat{U}_n$.
The depth of these circuits scales at worst $\mathcal{O}(mK^4)$ where $m$ is the trotter number and $K$ is the number of molecular orbitals.
At the minimal Trotter number level $(m=1)$, the maximum circuit depth for \methodabbr is comparable to that of UCC with generalized singles and doubles (employng the same Trotter number), and far shallower than QPE.
More importantly, the circuit depth of \methodabbr is independent of size of the Krylov basis one wishes to generate, allowing for a flexible trade off-between quantum and classical cost (for a desired level of accuracy). 
For example, in the NISQ device era, one may avoid larger circuit depths with \methodabbr by employing a modest Trotter number, but still achieve a high degree of accuracy by building a larger Krylov space that will be diagonalized classically.
In this way \methodabbr has both the advantage of selected CI to exploit wave function sparsity and the classical compression afforded by its quantum computational subroutines.
This flexibility is a feature that distinguishes \methodabbr from other QSD methods.

\begin{table}[!bh]
\centering
\renewcommand{\arraystretch}{0.9}
\caption{Ground-state energies (in \Eh) of \ce{H6} and \ce{H8} at a site-site distance of 1.5~{\AA} using exact time-evolution. Energy and overlap condition number $k(\mathbf{S})$ results are given for a single determinant (QK) using $N$ Krylov basis states and $\Delta t = 0.5$.
\methodabbr results are given for $N = d (s+1)$ Krylov basis states using three steps ($s = 3$) and $\Delta t = 0.5$ a.u.
With $N$ greater than 12 states, the condition number for QK does not grow larger than $10^{18}$.
This is likely a result of limitations of double precision arithmetic.
}
\footnotesize 
\begin{tabular*}{1.0\columnwidth}{@{\extracolsep{\stretch{1.0}}}*{1}{r}*{4}{r}@{}}
    \hline

    \hline
     $N$  & $E_{\rm{QK}}$ & $k(\mathbf{S}_{\rm{QK}})$  & $E_{\rm{\methodabbr}}$& $k(\mathbf{S}_{\rm{\methodabbr}})$    \\
    \hline
  \multicolumn{5}{c}{\ce{H6} ($r_\mathrm{HH}$ = 1.5 \AA{})}\\
  4 & $-$3.015510 & 3.29$\times 10^{5}$ & $-$3.015510 & 3.29$\times 10^{5}$  \\
  8 & $-$3.019768 & 3.60$\times 10^{11}$ & $-$3.019301 & 4.86$\times 10^{5}$  \\
 12 & $-$3.020172 & 1.61$\times 10^{17}$ & $-$3.019696 & 9.39$\times 10^{5}$ \\
 16 & $-$3.020192 & 3.19$\times 10^{17}$ & $-$3.019835 & 5.68$\times 10^{6}$ \\
 20 & $-$3.020198 & 3.86$\times 10^{17}$ & $-$3.019929 & 6.23$\times 10^{6}$ \\[6pt]
		~~~FCI       &  $-$3.020198          \\[8pt]
  \multicolumn{5}{c}{\ce{H8} ($r_\mathrm{HH}$ = 1.5 \AA{})}\\
  4 & $-$4.017108 & 1.19$\times 10^{5}$ & $-$4.017108 & 1.19$\times 10^{5}$  \\
  8 & $-$4.026563 & 1.39$\times 10^{10}$ & $-$4.024268 & 1.50$\times 10^{5}$ \\
12 & $-$4.028000 & 5.11$\times 10^{14}$ & $-$4.025894 & 2.00$\times 10^{5}$  \\
16 & $-$4.028096 & 1.33$\times 10^{17}$ & $-$4.026042 & 2.51$\times 10^{5}$  \\
 20  &  -- \quad & -- &$-$4.026387 & 4.27$\times 10^{5}$ \\
 24 &  -- & -- &$-$4.026457 & 4.44$\times 10^{5}$ \\[6pt]
 		~~~FCI       &  $-$4.028152          \\
    \hline
    
    \hline
\end{tabular*}
\label{tab:Econv_comparison}
\end{table}

\section{Computational Details}
The \methodabbr method was implemented using both an exact second quantization formalism and a quantum computer simulator using the open-source package \textsc{QForte}.\cite{Evangelista2019QForte} 
All calculations used restricted Hartree--Fock (RHF) orbitals generated with \textsc{Psi4}\cite{Parrish2017Psi4} using a minimal (STO-6G)\cite{Hehre1969ASelf} basis.
Molecular Hamiltonians for the hydrogen and \ce{BeH2} systems were translated to a qubit representation via the Jordan--Wigner transformation as implemented in \textsc{OpenFermion}\cite{McClean:2017tj} with default term ordering.
For all calculations, references in \methodabbr were selected using initial QK calculations with $s_0 = 2$ evolutions of the Hartree--Fock determinant and a time step of $\Delta t = 0.25$ a.u.   
Parameters such as the time time step ($\Delta t$), and number of evolutions per reference ($s$) used in \methodabbr were chosen based on energy accuracy and numerical stability.
We also note that we take the Trotter approximation with $m=100$ as a good approximation to the infinite $m$ limit for the potential energy curves we plot.
Adaptive derivative-assembled pseudo-Trotter ansatz variational quantum eigensolver (ADAPT-VQE)\cite{Grimsley:2019ed} calculations were performed with a in-house code provided by N. Mayhall.

\section{Numerical Studies and Discussion}
\label{sec:results}

We benchmark the performance and comparative numerical stability of the \methodabbr algorithm with linear chains of six and eight hydrogen atoms, two canonical models for one-dimensional materials with correlation strength modulated by bond length.\cite{Motta2017TowardsThe, Sinitskiy2010StrongCorrelation, stella2011strong}
We utilize point-group symmetry, which results in a determinant space comprised of 200 and 2468 determinants for \ce{H6} and \ce{H8}, respectively.
We first consider \ce{H6} at a site-site distance of 1.50~{\AA}, which exhibits strong electron correlation, as indicated by the large correlation energy ($E_{\rm{corr}}= -0.24681$~\Eh) and the small weight of the Hartree--Fock determinant in the FCI expansion ($|C_{\rm{HF}}|^2=0.634$).

In Tab.~\ref{tab:Econv_comparison} we show a comparison of the energy and overlap matrix condition number for the single reference version of quantum Krylov (QK), taking only the HF determinant as a reference, and \methodabbr as a function of the total number of basis states.
For \ce{H6} we observe that in both the single and multireference cases, convergence to chemical accuracy (error less than 1 kcal mol$^{-1}$ = 1.594 m\Eh) is achieved with only 8 parameters, an order of magnitude smaller than the size of FCI space.
For the case $N = 12$, \methodabbr identifies the following three references
\begin{equation}
\begin{split}
\ket{\Phi_0} = & \ket{220200} \\
\ket{\Phi_1} = & \ket{200220} \\
\ket{\Phi_2} = &
-0.302 \ket{2\uparrow\uparrow\downarrow\downarrow0} 
-0.302 \ket{2\downarrow\downarrow\uparrow\uparrow0} \\
& +0.275 \ket{2\uparrow\downarrow\uparrow\downarrow0} 
+0.577 \ket{2\uparrow\downarrow\downarrow\uparrow0} \\
&+0.577 \ket{2\downarrow\uparrow\uparrow\downarrow0}
+0.275 \ket{2\downarrow\uparrow\downarrow\uparrow0}
\end{split}
\end{equation}
where the orbitals are ordered according to ($1a_g$, $2a_g$, $3a_g$, $1b_{1u}$, $2b_{1u}$, $3b_{1u}$) in the D$_{2h}$ point group. These references are comprised of two closed-shell and six open-shell determinants. If we perform a computation with a set of references consisting of eight individual (uncontracted) determinants, the resulting Krylov space has dimension 32 and the corresponding energy is $-$3.019797~\Eh, which is only 0.1~m\Eh lower than the contracted result ($-$3.019696).
Turning to \ce{H8}, we find that the single-reference QK energy converges to chemical accuracy with only 12 parameters, two orders of magnitude fewer than FCI.
For the same example, the \methodabbr energy error is 1.06 kcal mol$^{-1}$ with 24 parameters, only slightly higher than chemical accuracy.

The linear dependency of the basis for \ce{H6} and \ce{H8}---as measured by the condition number of the overlap matrix [$k(\mathbf{S})$]---is significantly more pronounced in the single reference QK than the \methodabbr version.
In the case of \ce{H6},  even with a small Krylov basis (8 elements), QK is potentially ill-conditioned [$k(\mathbf{S}) =  3.60\times 10^{11}$].
In the case of 12 (or more) states, the QK eigenvalue problem is strongly ill-conditioned [$k(\mathbf{S}) =  1.16\times 10^{17}$], while \methodabbr displays only a modest condition number,  [$k(\mathbf{S}) =  9.39\times 10^{5}$].
Importantly, QK becomes ill-conditioned before reaching chemical accuracy, whereas \methodabbr does not, highlighting the importance of multireference approach for practical applications.

\begin{table*}[!ht]
\centering
\renewcommand{\arraystretch}{0.9}
\caption{Ground-state energies (in \Eh) of \ce{H6} at a bond distance of 1.5~\AA{}. \methodabbr results are given for $N = d (s+1)$ Krylov basis states using three steps ($s = 3$) and $\Delta t = 0.5$ a.u.
The quantity $m$ indicates the Trotter number. For each value of $N$, selected configuration interaction (sCI) results were obtained using $N$ determinants with the largest absolute coefficient in the FCI wave function. ADAPT-VQE results show the energy with $N$ cluster amplitudes selected from the pool of spin-adapted generalized singles/doubles.}
\footnotesize
\begin{tabular*}{2\columnwidth}{@{\extracolsep{\stretch{1.0}}}*{1}{r}*{7}{r}@{}}
    \hline

    \hline
     $N$  & $E^{(m=\infty)}_{\rm{\methodabbr}}$ & $E^{(m=8)}_{\rm{\methodabbr}}$ & $E^{(m=4)}_{\rm{\methodabbr}}$ & $E^{(m=2)}_{\rm{\methodabbr}}$ &  $E^{(m=1)}_{\rm{\methodabbr}}$  & $E_{\rm{sCI}}$ &  $E_{\rm{ADAPT-VQE}}$  \\
    \hline
		4          &  $-$3.015510	  & $-$3.014138    & $-$3.009948 	& $-$2.998858  &   $-$2.982186  & $-$2.845002 & $-$2.906724	 \\
		8   	   &   $-$3.019301 	  & $-$3.018341    & $-$3.015872 	& $-$3.010035  &   $-$3.001195  & $-$2.909404  & $-$2.983042   \\
		12   	   &   $-$3.019696 	  & $-$3.018808    & $-$3.016940	& $-$3.013425  &   $-$3.008661  & $-$2.926337 & $-$2.995691  \\
		16	   &   $-$3.019835	  & $-$3.018888    & $-$3.017173 	& $-$3.014253 &    $-$3.010543   & $-$2.954587 & $-$3.002345    \\
		20	   &   $-$3.019929	  & $-$3.019054    & $-$3.017614 	& $-$3.015311 &    $-$3.011663  & $-$2.961772 & $-$3.008847    \\[3pt]
		~~~FCI       &  $-$3.020198          \\
    \hline
    
    \hline
\end{tabular*}
\label{tab:Econv_comparison_trot_num}
\end{table*}

Next, we assess the errors introduced by approximating the real-time dynamics with a Trotter approximation.
Table~\ref{tab:Econv_comparison_trot_num} shows the performance of \methodabbr  using various levels of Trotter approximation for \ce{H6} at a bond distance of 1.5 \AA{}.
While using exact time evolution affords the fastest energy convergence with respect to number of Krylov basis states, we find that chemical accuracy can still be achieved using a Trotterized exponential.
For example, using a Trotter number $m=8$, \methodabbr gives an error of only 1.1~m\Eh with a basis of 20 Krylov states. 
In Table~\ref{tab:Econv_comparison_trot_num} we also show a comparison of \methodabbr with selected configuration interaction (sCI) and the adaptive derivative-assembled pseudo-Trotter ansatz variational quantum eigensolver (ADAPT-VQE).\cite{Grimsley:2019ed}
For any Trotter number, \methodabbr converges significantly faster than sCI and the ADAPT-VQE method.
For example, even with the smallest Trotter number ($m=1$) \methodabbr with 20 Krylov states gives an error of 8.5~m\Eh, while a sCI wave function with 20 determinants yields an error of 58.4~m\Eh (see Table~\eqref{tab:Econv_comparison_trot_num} for details of the determinant selection).
In comparison, an ADAPT-VQE wave function with 20 parameters yields an error of 11.4~m\Eh.
These results demonstrate the ability of \methodabbr to parameterize strongly correlated states efficiently using a small fraction of the variational degrees of freedom.

\begin{figure}[h!]
\centering
\includegraphics[width=3.2in]{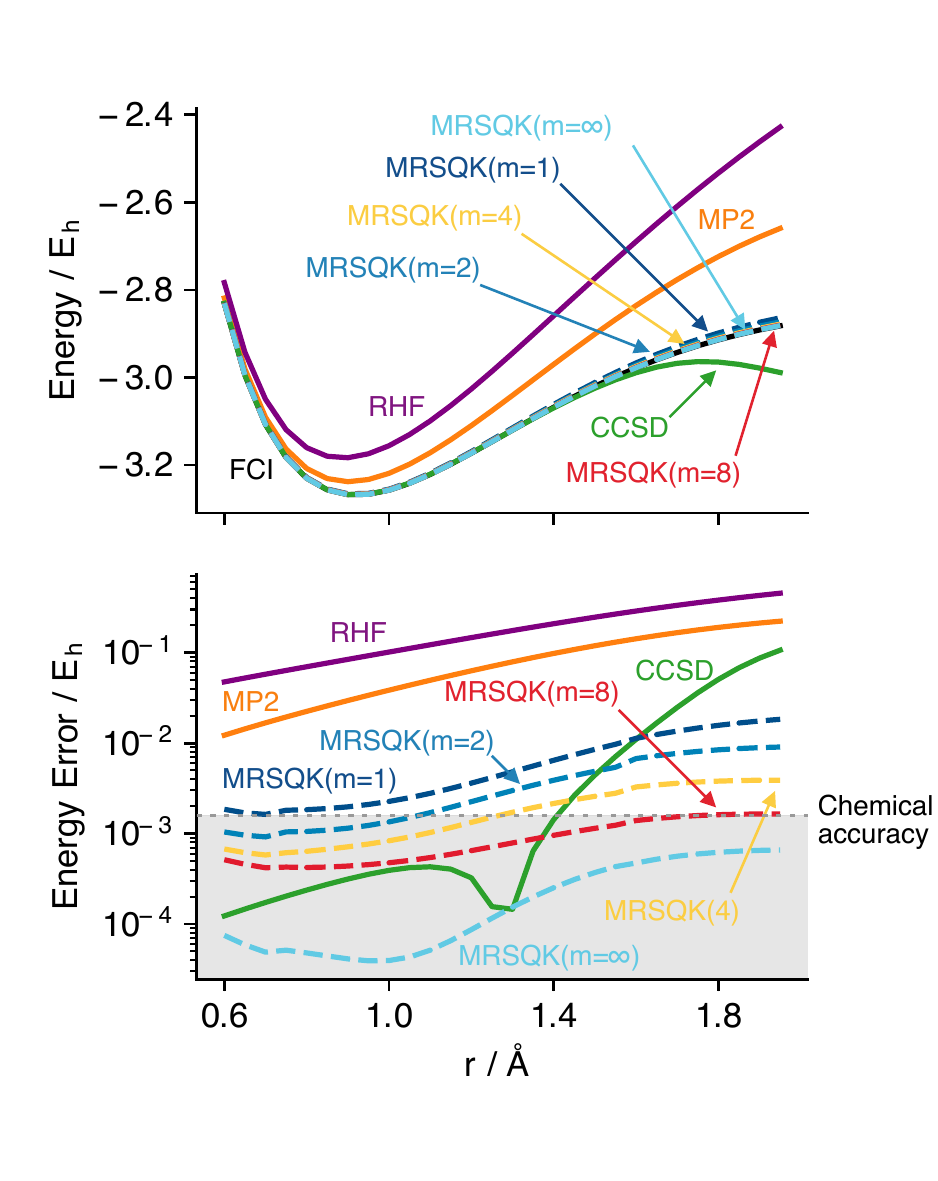}
\caption{Potential energy curve (top) and error (bottom) for symmetric dissociation of linear \ce{H6} in a STO-6G basis.
\methodabbr computations use $ \Delta t = 0.5$ a.u., three time steps ($s=3$), and five references ($d=5$) corresponding to 20 Krylov basis states. The number of Trotter steps ($m$) is indicated in parentheses, while those from exact time evolution are labeled ($m = \infty$).}
\label{fig:H6pes_and_error_5e_1}
\end{figure}
To illustrate the ability of \methodabbr to determine accurate ground-state potential energy surfaces (PES) in the presence of strong correlation, we examine the dissociation of the \ce{H6} chain and linear \ce{BeH2}.
Figure~\ref{fig:H6pes_and_error_5e_1} show the energy and error with respect to FCI for \ce{H6}, for restricted Hartree--Fock (RHF), second-order M{\o}ller--Plesset perturbation theory (MP2), coupled cluster with singles and doubles (CCSD),\cite{shavitt2009many} and \methodabbr with a Krylov basis of 20 states ($s=3$, $d=5$).
With the onset of strong electron correlation, single-reference methods (RHF, MP2, CCSD) fail to capture the the correct qualitative features of the PEC.
For example, CCSD produces very accurate results near the equilibrium geometry; however, it dips significantly below the FCI energy for bond distances greater than 1.5~\AA{}.
In contrast, \methodabbr far outperforms CCSD even with the lowest Trotter number ($m=1$) and chemically accurate \methodabbr results are obtained with $m=8$.

\begin{figure}[h!]
\centering
\includegraphics[width=3.2in]{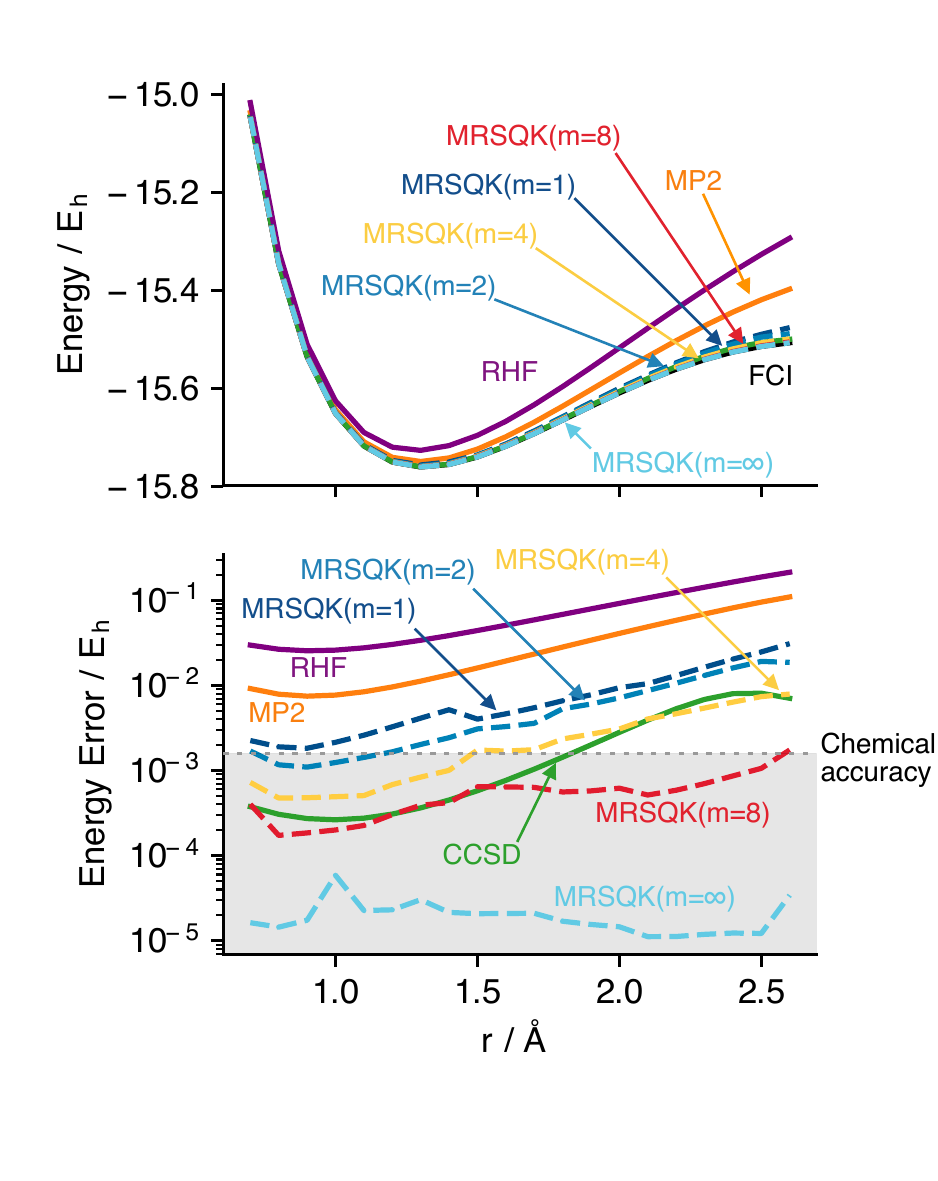}
\caption{Potential energy curve (top) and error (bottom) for symmetric dissociation of linear \ce{BeH2} in a STO-6G basis.
\methodabbr computations use $ \Delta t = 2$ a.u., four time steps ($s=4$), and six references ($d=6$) corresponding to 30 Krylov basis states. The number of Trotter steps ($m$) is indicated in parentheses, while those from exact time evolution are labeled ($m = \infty$).}
\label{fig:BeH2pes_and_error_5e_1}
\end{figure}

In Figure~\ref{fig:BeH2pes_and_error_5e_1} we report the potential energy curve for the symmetric dissociation of linear \ce{BeH2}. For this problem, the size of the determinant space is 169.
Like \ce{H6}, \ce{BeH2} is a challenging problem for single-reference methods, although CCSD shows smaller errors (less than 10 m\Eh) throughout the entire curve.
\methodabbr computations on \ce{BeH2} employed 30 Krylov states generated by a space of six references and four time steps ($s = 4$).
For this problem, we found that using a larger time step provides more accurate results and therefore, we report results using $\Delta t = 2$ a.u.
In the case of no Trotter approximation ($m = \infty$), the \methodabbr error is less than 0.1 m\Eh across the entire potential energy curve.
The approximate \methodabbr scheme based on four Trotter steps is already comparable in accuracy to CCSD, while using $m=8$ the error falls within chemical accuracy.
By analyzing the error plot in the bottom half of Fig.~\ref{fig:BeH2pes_and_error_5e_1}, we see that there are small discontinuities in the curve due to the selection of a different set of reference states.
This problem, however, is common to all selected CI methodologies, \cite{Huron:1973cb,  Schriber2016Adaptive, Holmes2016HeatBath, Tubman:2016jq,Garniron:2017ix} as well as ADAPT-VQE.
These discontinuities may be removed by employing references built from a fixed set of determinants.

\section{Conclusions}
In summary, the \methodname is a new quantum subspace diagonalization algorithm for solving the electronic Schr\"{o}dinger equation on NISQ devices.
\methodabbr diagonalizes the Hamiltonian in a basis of many-body states generated by real-time evolution of a set of orthogonal reference states.
This approach has two major advantages: (i) it requires no variational optimization of classical parameters, (ii) it avoids the linear dependency problem that may plague other \familyabbr methods.
Benchmark computations on \ce{H6}, \ce{H8}, and \ce{BeH2} show that \methodabbr with exact time-propagation converges rapidly to the exact energy using a number of Krylov states that is a small fraction of the full determinant space.
When the real-time propagator is approximated via a Trotter decomposition, modest Trotter numbers $m = 4,8$ are sufficient to ensure that truncation errors yield chemically accurate potential energy curves.
We also report a comparison of the convergence of the energy of \ce{H6} for \methodabbr,  selected configuration interaction (sCI), and the state-of-the-art ADAPT-VQE algorithm.
In comparing sCI and \methodabbr, the significantly faster convergence of the latter method indicates that the Krylov basis efficiently captures the important multideterminantal features of the wave function.
The comparison with ADAPT-VQE shows that \methodabbr can achieve a compact representation of the wave function competitive even with an adaptive strategy that aims to minimize the number of unitary rotations.

Together, these advantages make \methodabbr a promising tool for treating strongly correlated electronic systems with quantum computation.
However, there are several aspects of the \methodabbr that deserve more consideration.
The current reference selection strategy may produce different sets of references as the molecular geometry is changed, which in turn causes small discontinuities in potential energy curves. Selection procedures that, e.g., identify references from a small fixed set of orbitals could be used to address this issues.
In this work, we have selected fixed values for the time steps $t_n$. Schemes in which the time steps are treated as variational parameters may be able to represent states with a fewer number of Krylov states and are worth exploring.
Another important aspect is improving the approximation to the real-time dynamics. Our results indicate that low Trotter number approximations ($m = 1,2$) commonly used in other context introduce errors that are too large.
It would be desirable to explore the implementation of real-time dynamics via alternative methods, e.g. truncated Taylor series.\cite{Berry:2015ir}
An interesting alternative is to follow the strategy of Ref.~\citenum{Poulin:2018fb}, which employs an unphysical dynamics generated by a simple function of the Hamiltonian. This dynamics still spans the classical Krylov space and may be implemented with the same number of gates as a single Trotter number approximation.

\section*{Acknowledgements}
The authors thank Dr. Mario Motta and Dr. Nathan Wiebe for helpful discussions.
This work was supported by the U.S. Department of Energy under Award No.  DE-SC0019374.
N.H.S. was supported by a fellowship from The Molecular Sciences Software Institute under NSF grant ACI-1547580.



\bibliographystyle{achemso}
\bibliography{rtlrefs-Renke.bib,rtlrefs-Nick.bib,rtlrefs-Francesco.bib,extra.bib}{}
\end{document}